\documentclass[letter]{aa}

\usepackage{txfonts}
\usepackage{xspace}
\usepackage{natbib}
\usepackage{graphicx}

\newcommand{\aof}{A\,0535+26\xspace}

\newcommand{\exo}{EXO\,2030+375\xspace}
\newcommand{\twos}{2S\,1845-024\xspace}
\newcommand{\augsept}{August/September 2005\xspace}
\newcommand{\inte}{\textsl{INTEGRAL}\xspace}
\newcommand{\xte}{\textsl{RXTE}\xspace}

\newcommand{\pca}{\textsl{PCA}\xspace}
\newcommand{\hexte}{\textsl{HEXTE}\xspace}
\newcommand{\suzaku}{\textsl{Suzaku}\xspace}

\newcommand{\swift}{\textsl{Swift}\xspace}
\newcommand{\bat}{\textsl{BAT}\xspace}
\newcommand{\xspec}{\textsl{XSPEC v11}\xspace}
\newcommand{\ftools}{\textsl{FTOOLS 6.3.1}\xspace}
\newcommand{\kev}{\mathrm{keV}\xspace}
\newcommand{\ergs}{\mathrm{erg\,s^{-1}}\xspace}
\newcommand{\kpc}{\mathrm{kpc}\xspace}
\newcommand{\s}{\mathrm{s}\xspace}
\newcommand{\ks}{\mathrm{ks}\xspace}

\newcommand{\hzs}{\mathrm{Hz\,s^{-1}}\xspace}
\newcommand{\sps}{\mathrm{s\,s^{-1}}\xspace}
\newcommand{\G}{\mathrm{G}\xspace}
\newcommand{\chisq}{\chi^{2}_{\mathrm{red}}/\mathrm{dof}\xspace}

\begin{document}

\title{The pre-outburst flare of the \\
\aof \augsept outburst}

\author{I.~Caballero\inst{1} 
\and A.~Santangelo\inst{1} 
\and P.~Kretschmar\inst{2}
\and R.~Staubert\inst{1} 
\and K.~Postnov\inst{1,3}
\and D.~Klochkov\inst{1}
\and A.~Camero-Arranz\inst{4} 
\and M.~H.~Finger\inst{5}
\and I.~Kreykenbohm\inst{1,6} 
\and K.~Pottschmidt\inst{7,8} 
\and R.~E.~Rothschild \inst{9} 
\and S.~Suchy \inst{9} 
\and J~.Wilms\inst{10}
\and C.~A.~Wilson\inst{5}}
\offprints{I.~Caballero, \\
\email{isabel@astro.uni-tuebingen.de}} 
\institute{ Institut f\"ur Astronomie und Astrophysik, 
Sand 1, 72076 T\"ubingen, Germany 
\and ISOC, European Space Astronomy Centre
(ESAC), P.O. Box 78, 28691 Villanueva de la Ca\~{n}ada (Madrid), Spain  
\and Sternberg Astronomical Institute, 119999, Moscow, Russia 
\and GACE, Instituto de Ciencias de los Materiales, Universidad de Valencia, 
PO Box 20085, 46071 Valencia, Spain 
\and National Space Science and Technology
Center, 320 Sparkman Drive NW, Huntsville, AL, 35805, USA 
\and ISDC, 16 Ch.\ d'\'Ecogia, 1290 Versoix, Switzerland 
\and CRESST, University of Maryland Baltimore County, 1000 Hilltop Circle, 
Baltimore, MD 21250, USA
\and NASA Goddard Space Flight Center, Astrophysics Science Division, 
Code 661, Greenbelt, MD 20771, USA
\and CASS, University of California at San Diego, La Jolla, CA 92093-0424, USA  
\and Dr. Remeis-Sternwarte, Astronomisches Institut der Universit\"at 
Erlangen-N\"urnberg, Sternwartstr. 7, 96049 Bamberg, Germany   }

\date{Received $<$date$>$; Accepted $<$date$>$ }

\titlerunning{Pre-outburst flare of \aof}
\abstract{} {We study
the spectral and temporal behavior of the High Mass X-ray 
Binary \aof
during a `pre-outburst flare' which took place 
$\sim5\,\mathrm{d}$  before the peak 
of a normal (type I) outburst in \augsept. 
We compare the studied
behavior with that
observed during the outburst.}
{We analyse \xte observations that monitored \aof 
during the outburst. We complete spectral and timing analyses of the data.
We study the evolution of the pulse period, present energy-dependent 
pulse profiles both at the initial pre-outburst flare and close to 
outburst maximum, and measure how the cyclotron resonance-scattering 
feature (hereafter CRSF) evolves.} 
{We present three main results: a constant 
period $P$=$103.3960(5)\,\s$ is measured until periastron passage, 
followed by a spin-up with a decreasing period derivative of 
$\dot{P}$=$(-1.69\,$$\pm$${0.04})\,\times10^{-8}\sps $ at 
MJD 53618, and $P$ remains constant again at the end of the 
main outburst. 
The spin-up provides evidence for the existence of 
an accretion disk during the normal outburst.
We measure a CRSF energy of   
$E_{\mathrm{cyc}}$$\sim$$50\,\kev$
during the pre-outburst flare, and 
$E_{\mathrm{cyc}}$$\sim$$46\,\kev$ during the main outburst.
The pulse shape, which varies significantly during both pre-outburst 
flare and main outburst, evolves strongly with photon energy.
}{}


\maketitle

\section{Introduction}\label{sect:intro}

The Be/X-ray binary \aof \footnote{Referred to as 1A 0535+262 in 
SIMBAD} is a transient source, characterized by quiescent states 
with X-ray luminosity $L_{\mathrm{X}}\,$$\lesssim$$10^{36}\,\ergs$, 
interrupted by normal (type I) outbursts, associated in general
with periaston passage, when a luminosity 
$L_{\mathrm{X}}$$\sim$$10^{36-37}\,\ergs$ is reached,  
and giant (type II) 
outbursts for which $L_{\mathrm{X}}$$>$$10^{37}\,\ergs$. \aof was 
discovered by \citet{rosenberg75} during a giant outburst of 
luminosity level of $L_{(3-7\kev)}$$\sim1.2$$\times10^{37}\ergs$. 
Since then, five giant outbursts have been detected in 
October 1980 \citep{nagase82}, June 1983 \citep{sembay90}, 
March/April 1989 \citep{makino89}, February 1994 
\citep{finger94_2}, and May/June 2005 \citep{tueller05}.
Following the giant outburst in May/June 2005, 
two normal outbursts occurred in August/September 2005 \citep{finger05_1} 
and December 2005 \citep{finger05_2}.

The source \aof is in a binary orbit, with an O9.7IIIe optical companion 
HDE245770 \citep{giangrande80}, of orbital period 
$P_{orb}$$=$$110.3\,$$\pm$$0.3\,\mathrm{days}$ and eccentricity 
$e$=$0.47\,$$\pm$$0.02\,$ \citep{finger94_2}. 
The estimated distance of the system is $2\,\kpc$ 
\citep{steele98}. The measured pulse period of the neutron 
star at MJD 53614.5137 is $P$=$103.39315(5)\,\s$ \citep{caballero07}. 
During quiescence a spin-down trend of the neutron star has been 
reported \citep{finger96,hill07}, and
during giant outbursts the neutron star shows a strong
spin-up. In the June 1983 giant outburst, a spin-up of 
$\dot{\nu}$$\sim$$0.6\times10^{-11}\,\hzs$ was measured 
\citep{sembay90}. 
In the February 1994 giant outburst the spin-up reached 
$\dot{\nu}$$\sim$$1.2\times10^{-11}\,\hzs$ and quasi periodic 
oscillations were present, confirming the presence 
of an accretion disk \citep{finger96}.

The X-ray spectrum of \aof is typically described by a phenomenological 
model consisting of a power law with an exponential cutoff. 
Two absorption like features, interpreted as cyclotron resonance 
scattering features CRSFs (fundamental and first harmonic), 
have been observed in the spectrum 
at $\sim$$45\,\kev$ and $\sim$$100\,\kev$  \citep{kend94,grove95}. 
Recent observations during the \augsept normal outburst
with \inte and \xte \citep{kretschmar05,wilson05,caballero07}
and \suzaku \citep{terada06} have confirmed
this result. The centroid energy of the CRSF was measured 
during the main outburst at different luminosity levels 
and it was not found to vary with X-ray luminosity. 
 
In this Letter we focus on the \augsept normal outburst. 
At the onset of this outburst, a sharp, pre-outburst flare is
observed superimposed on the gradual increase towards the 
peak flux in the main outburst. This pre-outburst 
flare develops on a time scale of 1 day,
short compared to the 
three-week duration of the main outburst. As can be seen from 
the light curve measured by \swift/\bat , the flare is  
one of several such flares superimposed on the rising edge
of the main outburst (see \citealt{postnov08}).
The pre-outburst flare studied here reaches 
a PCA flux of $810\,\mathrm{counts\,s^{-1}\,PCU^{-1}}$, and
the maximum flux during the main outburst is 
$1005\,\mathrm{counts\,s^{-1}\,PCU^{-1}}$.  
The spectral and timing behaviours of the source
during the pre-outburst flare appear to be different from the ones
observed during the main outburst.

\section{Instruments and Observations}\label{sect:data}

\xte \citep{bradt93} observed \aof between 2005 August 28 and 
2005 September 24 in a target of opportunity observation completed  
as part of a campaign studying accreting pulsars. 
We present observations from 
the Proportional Counter Array \pca \citep[2--60keV,][]{jahoda96} 
and from the High Energy X-ray Timing Experiment \hexte 
\citep[20--200\,keV,][]{rothschild98}. 
A total of 44 pointed observations were completed to monitor 
the outburst with a total exposure time of $\sim140\,\mathrm{ks}$.
Fig.~\ref{fig:spin_up} 
shows the \pca light curve during the outburst 
in the $\sim$3--30$\,\kev$ range, obtained by averaging 
the individual light curves of all the observations. 
The analysis of \xte data was performed with \ftools 
and the spectral analysis with the X-Ray Spectral 
Fitting Package \xspec \citep{arnaud96}.


\section{Timing analysis}\label{sect:timing}
\subsection{Evolution of pulse period during the outburst}
\label{sect:period}

	\begin{figure}
	\resizebox{\hsize}{!}{\includegraphics{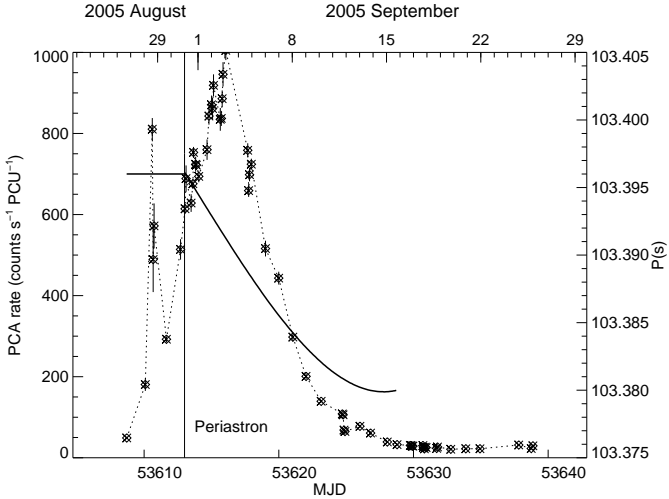}}
	\vspace{-6.mm}
	\caption{Left axis, dotted line: $\sim$3--30$\,\kev$ 
	  PCA mean count rate for each 
	  available observation
	  during the \augsept outburst. 
	  Each diamond indicates 
	  one \xte observation. The mean was obtained 
	  averaging each individual light curve. 
	  The errors 
	  shown are the sigma of the distribution for each observation.
	  Right axis, solid line: Pulse period evolution during the outburst. 
	  The vertical line indicates the time 
	  of periastron.
	  }\label{fig:spin_up}
	\end{figure}

We measured the pulse period of the pulsar with high 
accuracy during the outburst. 
For each available observation, we extracted  
barycentric-corrected \pca and \hexte light curves and corrected the photon 
arrival times for the orbital motion. We used the orbital ephemeris
from \citet{finger94_2} with the periastron epoch updated to
MJD 53613.0$\pm$1.3 \citep{finger06,finger07}.  
We then produced pulse profiles using a constant, trial period obtained 
using \inte observations of the same outburst \citep{caballero07}. 
For both \pca and \hexte we determined a reference 
time in the pulse profiles, 
and performed a phase-connection analysis
as previously completed in \citealt{caballero07} (see 
also \citealt{ferrigno07}). As a phase reference we selected the 
midpoint of a sharp edge present in the
pulse profiles, an apparently stable feature. 
The pulse period is approximately
constant during the initial pre-outburst flare, 
and a significant spin-up is measured after periastron. 
The period evolution after periastron can be described 
by a third-order polynomial. The results of the fit
are provided in Table~\ref{tab:period}. The period at any time
during the outburst is well-described using those functions, 
and $\pm\,5\,\mathrm{ms}$ is a conservative estimate of the 
uncertainty.   
The function describing the period evolution is plotted 
in Fig.~\ref{fig:spin_up}.
It is consistent with the period measured using 
\inte data within uncertainties \citep{caballero07}.

	\begin{table}
	\caption{Formal function decribing the  pulse period during the 
	outburst. The reference for the third order polinomial fit is MJD 
	53618. From MJD 53629 the period is essentially constant.}
	\label{tab:period} \centering
	\renewcommand{\arraystretch}{1.2}
	\begin{tabular}{ccc}\hline
&MJD& MJD\\
&$53608.70-53613.02$ &  $53613.11-53629$ \\\hline
$P(\s)$         & $103.3960(5)$        &$103.3883(5)$\\
$\dot{P}(\sps)$                      &-&$(-1.69\,\pm{0.04})\,\times10^{-8}$\\
$\ddot{P}(\mathrm{ss^{-2}})$     &-&$(9\,\pm{3})\,\times10^{-15}$\\
$d^{3}P/dt^{3}(\mathrm{ss^{-3}})$&-&$(2.5\pm{0.9})\times10^{-20}$ \\\hline
	\end{tabular}
  	\renewcommand{\arraystretch}{1.0}
	\end{table}

\subsection{Pulse profile evolution during the outburst}\label{sect:profiles}

\begin{figure}
	\resizebox{\hsize}{!}{\includegraphics{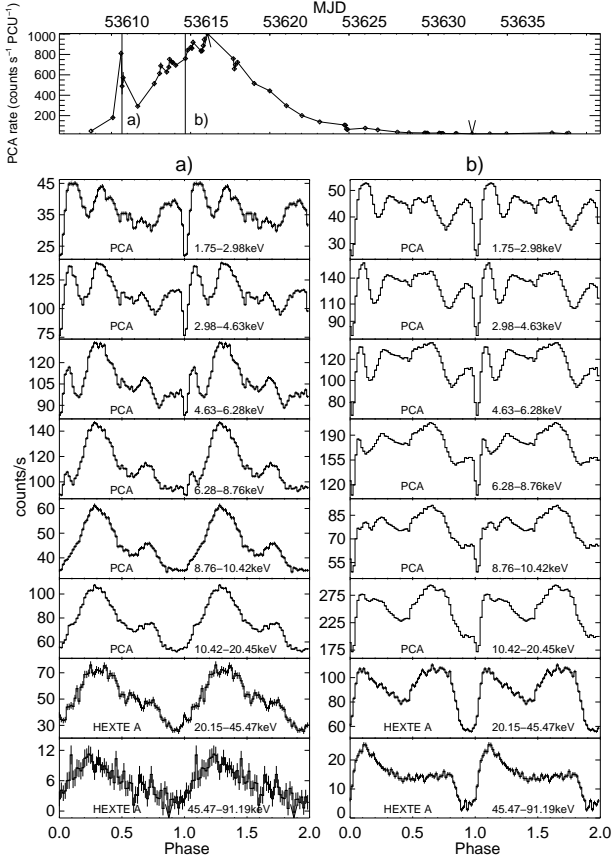}}
	\vspace{-12.mm}
	\caption{\pca and \hexte background substracted pulse profiles 
	from one observation during the pre-periastron flare (a), with 
	$L_{(5-50\kev)}$$\sim$$0.45\times10^{37}\ergs$, 
	and from one observation near the maximum of the outburst 
	(b), with $L_{(5-50\kev)}$$\sim$$0.75\times10^{37}\ergs$.
        Two pulse cycles are shown for clarity. 
	In the upper panel the PCA light curve is shown. The 
	vertical lines indicate the corresponding observations. 
        The exposure times are $2.2\,\ks$ in (a) and $12.35\,\ks$ 
	in (b).}\label{fig:pp}
	\end{figure}

Applying the measured period and its derivatives to 
fold light curves, we studied changes in 
the pulse profiles between the pre-outburst flare and 
the main outburst. 
Strong changes in the pulse-profile shape, with photon energy, 
are observed for both the main outburst and pre-outburst flare but 
in quite different ways.
Fig.~\ref{fig:pp} shows pulse profiles from observations (a), during the
pre-outburst flare, and (b), close to the maximum of the main
outburst. Observation (b) was chosen because of its high data 
quality and the pulse profiles are representative of data acquired during 
the outburst peak.
The low-energy pulse profiles in both cases show a complex
pattern, but different structures and evolution. At higher 
energies both pulse profiles show a simpler two peaked shape. 
During the main outburst, one of the peaks is strongly reduced
at the cyclotron energy. During the pre-outburst flare there is a smooth 
evolution towards higher energies without a change crossing the 
cyclotron energy. Energy-dependent pulse profiles for all the
observations during the outburst are shown in \citet{camero07} and
a detailed study will be presented in a forthcoming paper \citep{camero07_2}.

\section{Spectral analysis}\label{sect:spec}

To model the phase-averaged, X-ray continuum, we 
have used a phenomenological model, 
a power law times an exponential 
cut-off (\textsl{XSPEC} model {\tt{cutoffpl}}), 
which is the simplest capable of reproducing the data.
It is described by $ F(E)$$\propto$$E^{-\alpha}e^{-E/E_{\rm fold}}$,
where $\alpha$ is the photon index and $E_{\rm fold}$ the 
folding energy. One, and in some observations two, absorption 
features were included in the model to enable 
an accurate description of the data. We modeled them with a 
Gaussian optical depth 
profile,  which modifies the continuum in the 
following way: $F'(E)$=$F(E)e^{-\tau(E)}$, where
 $\tau(E)$=$\tau e^{-(E-E_{\mathrm{cyc}})^{2}/(2\sigma^{2})}$.
A Gaussian emission line was added to the model to account
for the Fe K$\alpha$ fluorescence line, with 
energy fixed at $6.4\kev$ and width to $0.5\,\kev$. 
Due to a feature in the residuals around $\sim4.7\,\kev$ 
(instrumental Xenon L edge, see \citealt{rothschild06}), 
we excluded data for energies  below $5\kev$ in our analysis. 

\begin{figure}[!h]
	\resizebox{\hsize}{!}{\includegraphics{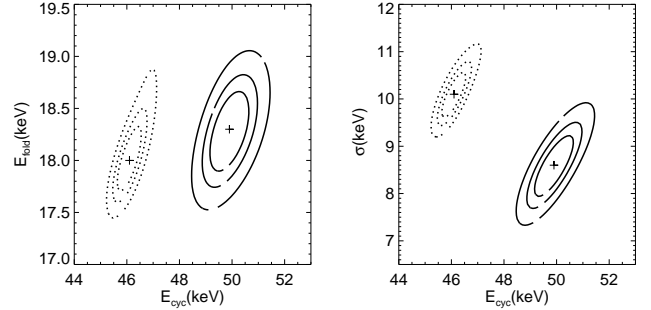}}
	\vspace{-4.mm}
	\caption{$E_{\mathrm{fold}}$ vs $E_{\mathrm{cyc}}$ (left) 
	and $\sigma$ vs $E_{\mathrm{cyc}}$ 
	(right)	contour plots for one observation near the 
	maximum (dotted lines, observation (b) in Fig.~\ref{fig:pp})
	and for the sum of the three available 
	observations during the pre-outburst flare (solid lines). 
	The contours indicate 
	$\chi^{2}_{\mathrm{min}}$+$2.30(68\%), 4.61(90\%), 9.21(99\%)$ levels.}
	\label{fig:contour}
	\end{figure}

We performed a spectral analysis of all observations available 
and studied the evolution of the cyclotron-line energy and continuum 
parameters during the outburst. We measured a constant value for 
the centroid energy of the fundamental CRSF
during the main outburst, 
as reported by \cite{caballero07}. 
During the pre-outburst flare, the cyclotron-centroid
energy is measured at a higher value, reaching 
$E_{\mathrm{cyc}}$=$52.0^{+1.6}_{-1.4}\,\kev$, compared to 
$E_{\mathrm{cyc}}$=$46.1^{+0.5}_{-0.5}\,\kev$ during the main outburst
(at 90\% confidence). 
To study the significance of the change,
we have produced $\chi^{2}$ contour plots 
for the observations during 
the flare and during the main outburst. As an example, 
in Fig.\ref{fig:contour} contour plots are shown for 
the observation close to the maximum of the main outburst 
labeled (b) in Fig.~\ref{fig:pp}, and for the sum of the three available
observations during the pre-outburst flare. In observation (b)
we included a harmonic, cyclotron line with the energy fixed
at $E_{\mathrm{cyc}}$=$102.5\,\kev$ (it is measured at 
$E_{\mathrm{cyc}}$=$102.5^{+4.5}_{-3.3}\kev$ at 90\% confidence). 
We conclude from the contour plots that the change in energy 
is statistically significant. 

Table~\ref{tab:spike_obs} contains the best-fit values 
for the three observations during the pre-outburst 
flare and for the sum of those observations, 
as well as the best-fit values for the main outburst  
(observation (b) in Fig.~\ref{fig:pp}). 
	
	\begin{figure}[h]
	\resizebox{\hsize}{!}{\includegraphics{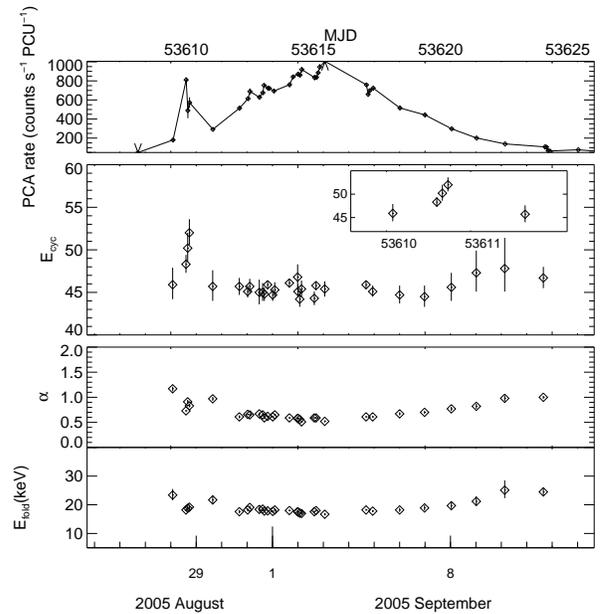}}
	\vspace{-4.mm}
	\caption{ First panel: PCA light curve. 
	Second panel: fundamental CRSF energy during the 
	ourbusrt. Inset shows a zoom on the 
	pre-outburst flare. Third and forth panels: photon index and folding
	energy evolution. Errors are $90\%$ confidence for one parameter of 
	interest ($\chi^{2}_{min}$+2.7).}\label{fig:cyc}
	\end{figure}
 	\begin{table*}[H]
	\centering
	\renewcommand{\arraystretch}{1.2}
	\caption{Best-fit values for the available observations 
	during the pre-outburst flare, where the CRSF energy is 
	measured at a higher position, for the sum of the observations 
	during the flare, and for one observation close to the maximum
	of the main outburst. L is the luminosity in 
	units of $10^{37}\ergs$ in the $5-50\,\kev$ range.
	Errors are $90\%$ confidence for one parameter of interest
	($\chi^{2}_{min}+2.7$).}
	\begin{tabular}{ccccccccccccc}\hline
 Obs. ID &$E_{\mathrm{cyc}}(\kev)$     & $\sigma(\kev)$ & $\tau$                 & $E_{\mathrm{fold}}(\kev)$& $\alpha$& MJD start & Exp. $(\ks)$& $L$&$\chisq$\\\hline
 91086-01-03-01  &$48.3^{+1.1}_{-1.0}$      & $6.4^{+1.0}_{-1.0}$  & $0.27^{+0.03}_{-0.03}$ & $18.2^{+0.4}_{-0.4}$& $0.73^{+0.02}_{-0.02}$&53610.599 &      1.152          & 0.85  & 1.19/170  \\
 91086-01-03-02  & $50.2^{+1.8}_{-1.6}$ & $8.7^{+1.4}_{-1.3}$  & $0.57^{+0.08}_{-0.08}$ & $18.8^{+1.0}_{-1.0}$ &$0.91^{+0.04}_{-0.04}$  &53610.665 &      2.208      & 0.45       & 0.88/170  \\
 91086-01-03-03  & $52.0^{+1.6}_{-1.4}$ & $10.8^{+1.3}_{-1.1}$ & $0.53^{+0.06}_{-0.06}$ &$19.2^{+0.8}_{-0.7}$ &$0.83^{+0.03}_{-0.03}$&53610.729 &     2.560         & 0.57   & 0.90/170\\
 sum             & $49.9^{+0.8}_{-0.8}$ & $8.6^{+0.7}_{-0.7}$&  $0.40^{+0.03}_{-0.03}$&$18.3^{+0.4}_{-0.4}$& $0.77^{+0.02}_{-0.02}$ &53610.599 &     5.920          &0.66   & 1.18/220  \\\hline
91085-01-01-03  & $46.1^{+0.5}_{-0.5}$ &  $10.1^{+0.5}_{-0.5}$&$0.47^{+0.03}_{-0.02}$&$18.0^{+0.4}_{-0.3}$&$0.59^{+0.02}_{-0.02}$ &53614.7& 12.352& 0.75& 1.19/214 \\
		 & 102.5 &  $8.4^{+2.2}_{-1.9}$ & $1.0^{+0.4}_{-0.2}$ (1st harmonic)\\\hline

	\end{tabular}
	\renewcommand{\arraystretch}{1.0}
	\label{tab:spike_obs}
	\end{table*}

Fig.~\ref{fig:cyc} shows the evolution of the CRSF energy during the
outburst, as well as the evolution of the continuum parameters
photon index $\alpha$ and folding energy $E_{\rm fold}$ for the observations
in which the cyclotron line was detected. The continuum parameters are
quite variable at the beginning of the outburst. The spectrum at the
beginning of the flare appears to be softer, with photon index 
$\alpha$$\sim$$1.2$, and then becomes harder in the flare. 
The rising of the main outburst presents a harder spectrum, 
with $\alpha$$\sim$$0.6$. The decay of the outburst shows a smooth  
softening of the spectrum.

 \section{Summary and Discussion}\label{sect:summ}

In this paper, we present evidence for significant changes 
in the spectral and timing parameters of the Be/X-ray binary 
\aof, during its normal outburst in \augsept. 
Our three main results are the following:
 \begin{enumerate}
 \item The pulse period of the neutron star 
appears to be constant during the 
pre-outburst flare, $P$=$103.3960(5)\,\s$, 
and a spin-up starts at periastron, 
$\dot{P}$=$(-1.69\,\pm{0.04})\,\times10^{-8}\sps$ 
measured at MJD 53618.
The pulse period falls exponentially at the end of the outburst.

\item The energy-dependent, pulse 
profiles during the pre-outburst flare are significantly different 
from those measured for the main outburst.

\item During the pre-outburst flare,
the fundamental cyclotron-line energy centroid 
reaches $E_{\mathrm{cyc}}$=$52.0^{+1.6}_{-1.4}\,\kev$, 
significantly higher than for the main outburst, 
$E_{\mathrm{cyc}}$=$46.1^{+0.5}_{-0.5}\,\kev$. 
\end{enumerate}

This is the first observation of a normal outburst that measures
a spin-up rate for \aof, providing evidence that an accretion disk 
is being detected during a Type I outburst.
The measured spin-up 
$\dot{P}$=$(-1.69\,\pm{0.04})\,\times10^{-8}\,\sps$, or 
$\dot{\nu}$=$(1.58\,\pm{0.04})\,\times10^{-12}\,\hzs$, 
is smaller than the spin-up measured during giant outbursts in the past, 
e.g., $\dot{\nu}$$\sim$0.6$\times10^{-11}\,\hzs$ in June 1983 
\citep{sembay90}, or $\dot{\nu}$$\sim$$1.2\times10^{-11}\,\hzs$ 
in February 1994 \citep{finger96}. 

The pulse profiles measured during the main 
outburst are consistent 
with those during a giant outburst 
in March/April 1989 albeit at a different luminosity 
level, $L_{(23-53\kev)}\,$$\sim$$1.26\,\times10^{37}\,\ergs$  
\citep{kend94}. A comparison between these
pulse profiles and those obtained for the pre-outburst flare 
in this paper suggest
that a different mode 
of accretion takes place during the flare 
(see \citealt{postnov08}).

The energy of the CRSF during the main outburst is 
consistent with the energy measured
with \inte close to the maximum of the main outburst,
$E_{\mathrm{cyc}}$=$45.9\pm0.3\,\kev$ \citep{caballero07} and with
the energy measured with \suzaku at 
the end of the main outburst, $E_{\mathrm{cyc}}$=$46.3^{+1.5}_{-1.3}\,\kev$
(using the same Gaussian model for the line, \citealt{terada06}).
From the observed cyclotron energy, 
the estimated, magnetic field at the site of the X-ray
emission during the main
outburst, assuming a redshift of 
z=0.3, is $B$$\sim$$5.2\times10^{12}\G$, and during the pre-outburst 
flare $B$$\sim$$5.8\times10^{12}\G$.
During the main part of the outburst the cyclotron-line energy 
remains constant within the errors, in spite of the changes in the
luminosity. 
This suggests that the structure of the accretion column is different 
from that observed in both V~0332 +53 \citep{mowlavi06} and 4U~0115+63 
\citep{tsygankov06}. The source is likely to be in the 
sub-Eddington regime, as for Her X-1 \citep{staubert07} for which 
it is believed  no shock has formed in the accretion column.

Pre-outburst flares have been observed in other accreting pulsars, 
such as \twos \citep{finger99} and \exo \citep{camero05}. 

\citet{hayasaki06} have modeled accretion disks around neutron stars in
Be/X-ray binaries. Their SPH simulations reproduce a series of normal
outbursts from the
accretion disk formed about the neutron star at periastron.  
In some cases, single flares are found preceding
the outburst maximum. We note first of all that the timescale of the flare
in their simulations appears to be too long (10\% of the orbital period)
compared to the
flare duration observed.
The inspection of the \swift-\bat
light-curve of A0535+26 during the outburst studied here shows that RXTE actually
observed only one of a collection of short flares in the rising part
of the outburst \citep{postnov08}. Such
flares are not reproduced by the above simulations. This model does not
take into account the disk-magnetospheric interaction.

A more plausible interpretation is that the pre-outburst 
flares could be caused by magnetospheric 
instabilities between the accretion disk and the
neutron-star magnetosphere at the onset of accretion. 
The instability causes the plasma that has accumulated 
in the disk-magnetosphere, 
boundary-layer to rapidly fall onto the neutron star 
surface close to the magnetic poles, but along different 
field lines than (quasi)stationary accretion 
(see \citealt{postnov08} for details).

\begin{acknowledgements}
We thank ISSI (Bern), 
for their hospitality during the team meetings held there.
IC thanks ESAC (ESA) in Madrid 
for their hospitality. This research is supported by the 
Bundesministerium f\"ur Wirtschaft und Technologie through the 
German Space Agency (DLR) under contract no. 50 OR 0302.
\end{acknowledgements}
\vspace*{-6.mm}
\bibliographystyle{aa}
\bibliography{references,a0535}

\end{document}